\documentclass[aps,prb,twocolumn,superscriptaddress,preprintnumbers,amsmath,amssymb,axodraw]{revtex4-2}

\usepackage{latexsym}
\usepackage{comment}
\usepackage{color}
\usepackage{amssymb}
\usepackage{mathtools}
\usepackage{amsmath}
\usepackage{bm}
\usepackage{amsfonts}
\usepackage{bbold}
\usepackage{cancel}
\usepackage{slashed}
\usepackage{appendix}
\usepackage{float}
\usepackage{cancel}
\usepackage{graphicx} 
\usepackage{dcolumn} 
\usepackage{braket}
\usepackage{tikz}
\usepackage{tikz-feynman}
\usepackage{axodraw2}
\usepackage{subcaption}
\usepackage{graphicx}
\usepackage[font={small}]{caption}
\usepackage{amsbsy}
\usepackage{orcidlink}
\usepackage{soul}

\newcommand{\bea}{\begin{eqnarray}}
\newcommand{\eea}{\end{eqnarray}}
\newcommand{\be}{\begin{equation}}
\newcommand{\ee}{\end{equation}}

\newcommand{\mbf}{\mathbf}

\begin{document}

\title{Phonon-based determination of elastic coefficients in the Weyl semimetal TaAs}
\author{Fabi\'an Jofr\'e-Parra}
\email{fljofre@uc.cl}
\affiliation{Facultad de F\'isica, Pontificia Universidad Cat\'olica de Chile, Vicu\~{n}a Mackenna 4860, Santiago, Chile}

\author{Debankita Ghosh}
\affiliation{Facultad de F\'isica, Pontificia Universidad Cat\'olica de Chile, Vicu\~{n}a Mackenna 4860, Santiago, Chile}

\author{Enrique Mu\~noz~\orcidlink{0000-0003-4457-0817}}
\email[Corresponding author: ]{ejmunozt@uc.cl}
\affiliation{Facultad de F\'isica, Pontificia Universidad Cat\'olica de Chile, Vicu\~{n}a Mackenna 4860, Santiago, Chile}
\affiliation{Center for Nanotechnology and Advanced Materials CIEN-UC, Avenida Vicuña Mackenna 4860, Santiago, Chile}

\date{\today}

\begin{abstract}
Reliable determination of elastic properties in topological semimetals is essential for understanding strain-related effects, but is often hindered by methodological and computational limitations. In this work, we combine first-principles phonon calculations with an elastic continuum model to determine the elastic properties of Weyl semimetal TaAs. The sound velocities extracted from the acoustic phonon branches are used to obtain the full set of elastic moduli, which show good agreement with previously reported values in the literature. From these results, we derive standard elastic parameters such as bulk, shear, and Young’s moduli, and the Poisson ratio. This approach highlights a computationally efficient alternative to conventional strain-based methods, avoiding the need for large supercells when shear is applied, and thus possible strain-induced inconsistencies in the electronic basis, while providing a possibility to connect with experimental characterizations (e.g. Raman or Brillouin-Mandelstam scattering) of the lattice dynamics in Weyl semimetals.
\end{abstract}

\maketitle 

\section{Introduction}

Transition-metal monopnictides have emerged in recent years as a paradigmatic class of topological quantum materials. Compounds such as TaAs, TaP, NbAs, and NbP host Weyl semimetal phases, characterized by pairs of Weyl nodes with opposite chirality acting as monopoles of Berry curvature in momentum space~\cite{Huang_2015, Xu_2015, Weng_2015, Lv_2015}. These materials exhibit topologically protected surface Fermi arcs and a variety of unconventional transport properties, including large magnetoresistance~\cite{Zhang_2017, Xu_2017, Sankar_2018}, chiral anomaly-induced negative magnetoresistance~\cite{Huang_2015-2, Zhang_2016}, and anomalous thermoelectric responses~\cite{Bonilla_2021, Bonilla_2022, Bonilla_2023, Bonilla_2024}. Spin-orbit coupling (SOC) plays a central role in shaping their electronic band structure: it gaps nodal lines protected by crystalline symmetries and stabilizes isolated Weyl points~\cite{Weng_2015, Yang_2018}, thereby controlling both the band topology and the low-energy excitations. As a result, transition-metal monopnictides provide a fertile platform to explore the interplay between crystal symmetry, pseudo-relativistic effects, and transport in solids.

Beyond their electronic band topology, the physical properties of Weyl semimetals are strongly influenced by lattice degrees of freedom. In this context, strain has emerged as a powerful tuning parameter, leading to the rapidly growing field of {\it{straintronics}} \cite{Miao_2021, Bandyopadhyay_2021, Yang_2024, Boland_2024}. Mechanical deformations can modify band dispersion~\cite{Zhou_2016, Cortijo_2016}, shift Weyl node positions~\cite{Schindler_2020}, induce pseudo-gauge fields~\cite{Cortijo_2015, Arjona_2018}, and alter phonon-mediated transport processes~\cite{Xu_2017-2, Rinkel_2019}. Understanding the elastic response of topological materials is therefore essential for both fundamental studies and potential device applications. Despite the considerable experimental studies~\cite{Besara_2016, Boller_1963, Furuseth_1965, Laliberte_2020, Saini_1964, Willerstrom1984} conducted on TaAs, direct experimental measurements of its elastic coefficients are, to the best of our knowledge, still lacking. This absence of experimental benchmarks makes it particularly important to explore alternative theoretical strategies to determine elastic properties with controlled accuracy.

In this work, we present a comprehensive study of the electronic, phonon, and elastic properties of TaAs based on first-principles calculations combined with an elastic continuum model. Our approach departs from the conventional strain-stress methodology commonly used to compute elastic constants within density functional theory (DFT) \cite{deJong_2015, Cao_2018, Kiely_2021}. Instead, we exploit the long-wavelength limit of the phonon dispersion and its connection to elastic wave propagation in anisotropic media. By extracting sound velocities along high-symmetry directions and combining them with a continuum elastic model, we obtain the full set of independent elastic moduli without the need to apply finite shear strains. This strategy circumvents well-known numerical difficulties associated with symmetry breaking, large supercell requirements, and Pulay stress corrections~\cite{Vesti_2023, Vanpoucke_2015, Ulian_2022}. Moreover, the direct experimental probe of elastic constants involves quasi-static deformation processes (e.g. the tensile test to measure the Young modulus), which entails technical difficulties~\cite{Bottani01012018} such that direct mechanical measurements have not yet been reported for TaAs and other Weyl semimetals. Alternatively, lattice dynamics and phonon spectrum can be experimentally probed by Raman scattering~\cite{Liu_2015,Gao_JAP_2007,Mishra_PRB_2001,Ramos_ACSNANO_2023,Weinstein_PhysRevB_1973}, particularly second-order Raman~\cite{Carles_PhysRevB_1980,Gao_JAP_2007,Carvalho_2017,Hildebrandt_2023,Iatsunskyi_Optik_2026,Mishra_PRB_2001,SOURISSEAU_1991,Weber_PRB_1993,Weinstein_PhysRevB_1973} for acoustic modes, field-angle dependence~\cite{Laliberte_2020}, Brillouin-Mandelstam scattering~\cite{Guzman_ACS_2022,Bottani01012018,Kargar_APL_2018,Kargar_NP_2021,Wright_APL_2024,Wright_APL_2025}, X-ray diffraction~\cite{Lee_PRL_2022} and other techniques~\cite{Bottani01012018,Rigg_2014}. In this context, our proposed method, based on the accurate determination of the speed
of sound (by DFPT) to infer the elastic moduli from an
analytical elastic model, then offers an alternative and
direct route to compare directly with the experimental lattice dynamics in TaAs~\cite{Laliberte_2020}.

The article is organized as follows. In Sec.~\ref{sec:methods}, we describe the computational methodology, including the details of the DFT calculations and the formulation of the elastic continuum model used to relate sound velocities to elastic coefficients. In Sec.~\ref{sec:results}, we present and discuss our results, starting with the electronic band structure and Fermi surface, followed by the phonon dispersion, sound velocities, and elastic moduli obtained using both approaches. Finally, in Sec.~\ref{sec:conclusion}, we summarize our main findings and highlight the advantages and limitations of the phonon-based elastic methodology.

\begin{table*}
    \centering
    \caption{Lattice parameters and mass density.}
    \begin{tabular}{cccccc} \hline \hline
         & \textbf{LDA} & \textbf{PBEsol} \quad & \textbf{PBE} \quad & \quad \textbf{Other expt. works} \cite{Besara_2016, Boller_1963, Furuseth_1965, Laliberte_2020, Saini_1964, Willerstrom1984} \quad & \quad \textbf{Other DFT works} \cite{Buckeridge_2016, Chang_2016, Grassano_2018, Huang_2015, Liu_2017, Naher_2021, Ullah_2023} \quad \\ 
        $a$ ($\mathring{\text{A}}$) & 3.407 & 
 $3.429$ & $3.458$ & $3.437 \pm 0.002$ & $3.45 \pm 0.02$  \\ 
        $c$ ($\mathring{\text{A}}$) & 11.516 & 
 $11.587$ & $11.703$ & $11.644 \pm 0.006$ & $11.67 \pm 0.06$  \\ 
        $\rho$ (g/cm${}^{3}$)& 11.603 & 12.477 & $12.147$ & $12.2 \pm 0.3$ & $12.07 \pm 0.58$  \\ \hline \hline
    \end{tabular}
    \label{tab:cell-parameters}
\end{table*}

\section{Methodology}   \label{sec:methods}

\subsection{Numerical calculations}

All our DFT calculations were performed using the QUANTUM ESPRESSO package~\cite{Giannozzi_2009, Giannozzi_2017, Giannozzi_2020}. The interaction between core and valence electrons was modeled with projector augmented-wave (PAW) pseudopotentials~\cite{Blochl_1994}, and exchange–correlation effects were treated within the Perdew–Burke–Ernzerhof (PBE) generalized gradient approximation (GGA)~\cite{Perdew_1996}. SOC was included in both electronic and subsequent phonon calculations to capture the relevant features of the band structures and to ensure that these effects were consistently retained in the derived material properties.

Unit cell relaxation and electronic band structure calculations were performed with a $100$~Ry plane wave cutoff and a $12 \times 12 \times 12$ Monkhorst–Pack~\cite{Monkhorst_1976} $k$-point mesh, which was sufficient to ensure total energy convergence within $10^{-4}$~eV. Unit cell relaxation was stopped once all stress components were less than $10^{-7}$~Ry~/~$a_B^3$ and all force components were below $5 \times 10^{-7}$~Ry~/~$a_B$, where $a_B$ is the Bohr radius.

Phonon frequencies were obtained by performing a self-consistent electronic calculation on a $10 \times 10 \times 10$ $k$-point mesh and subsequently using density functional perturbation theory (DFPT) \cite{Baroni_2001} including the SOC~\cite{Urru_PhysRevB.100.045115}, as implemented in the \textit{ph.x} module, on a $5 \times 5 \times 5$~$q$-point mesh. The sound velocities were extracted from the phonon spectrum by performing a linear fit to the acoustic branches in the vicinity of the $\Gamma$ point.

The elastic moduli $C_{11}$, $C_{33}$, $C_{12}$, and $C_{13}$, expressed in Voigt notation~\cite{Authier_2006}, were obtained numerically by performing linear fits of the stress response to controlled applications of the normal strain components $u_{xx}$ and $u_{zz}$ within our DFT calculations. In contrast, determining the shear moduli $C_{44}$ and $C_{66}$ requires applying finite shear strains $u_{xz}$ and $u_{xy}$, respectively. Implementing these shear deformations is considerably more challenging, as they break the crystal symmetries and require very large supercells in order to remain within the linear-elastic regime.

\begin{figure}
  \centering
  \begin{subfigure}{0.185\textwidth}
    \centering
    \includegraphics[scale=0.32]{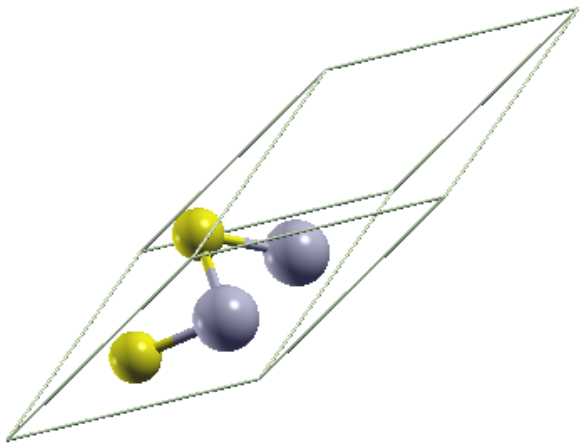}
    \label{fig:primitive-cell-TaAs}
  \end{subfigure}
  \begin{subfigure}{0.185\textwidth}
    \centering
    \includegraphics[scale=0.4]{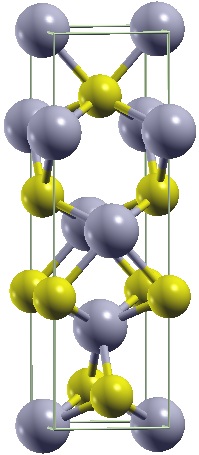}
    \label{fig:conventional-cell-TaAs}
  \end{subfigure}
\caption{Primitive (left) and conventional (right) unit cells. Ta and As atoms are shown as grey and yellow spheres, respectively.}
\label{fig:unit-cells-TaAs}
\end{figure}

\subsection{Elastic continuum model}

\begin{figure}[b]
    \centering
    \includegraphics[scale=0.48]{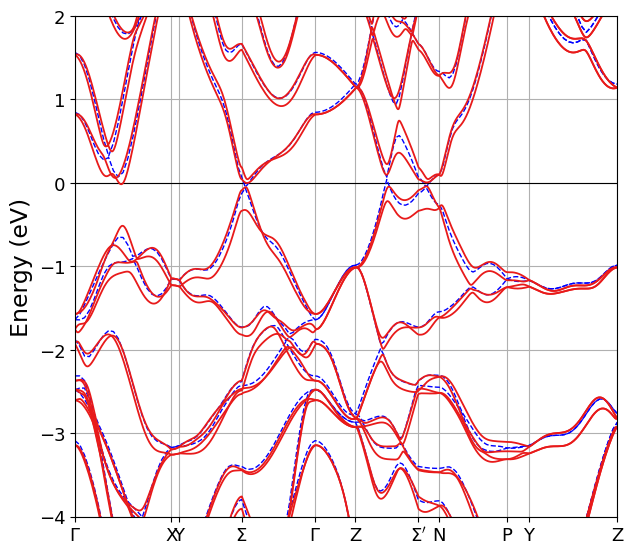}
    \caption{Electronic band structure computed with (red, solid line) and without (blue, dashed line) SOC.}
    \label{fig:bands}
\end{figure}

As a local constitutive property, the elasticity tensor is invariant under the crystallographic point group of TaAs, corresponding to a body-centered tetragonal lattice with $C_{4v}$ symmetry. This reduces the fourth-rank elasticity tensor to six independent coefficients~\cite{Authier_2006}. Starting from the stress–strain relation, we obtain the dynamical matrix by Fourier transforming the linearized equations of motion for the displacement field $\Delta\boldsymbol{x}(\mathbf{q},\omega)$, leading to $\left[ \mathbf{M}(\mathbf{q}) - \rho \omega^{2} \right]\Delta \boldsymbol{x}(\mathbf{q},\omega) = 0$, where $\rho$ is the mass density, $\omega$ is the elastic wave frequency, and $\mathbf{M}(\mathbf{q})$ denotes the dynamical matrix with momentum $\mathbf{q}$. The details of the derivation and diagonalization of $\mathbf{M}(\mathbf{q})$ are provided in Appendix~\ref{appx:sound-velocities}.

\begin{table*}
    \centering
    \caption{Optical phonon frequencies at the $\Gamma$ point.}
    \begin{tabular}{ccccc} \hline \hline
        & \textbf{PBEsol (cm${}^{-1}$)} & \quad \textbf{PBE} (cm${}^{-1}$) \quad & \quad \textbf{Other expt. work} \cite{Liu_2015} (cm${}^{-1}$) \quad & \quad \textbf{Other DFT works} \cite{Liu_2015, Chang_2016} (cm${}^{-1}$) \quad \\  
        $E(1)$ & 124.1 & 119.9 & 125.6 &  120.2  \\  
        $B_1(1)$ & 166.2  & 160.2 & 171.7 & 165.3  \\ 
        $E(2)$ & 232.2  & 223.7 & 232.6 & 223.4   \\ 
        $A_1$  & 251.2 & 243.8 & 251.9 & 242.6  \\  
        $E(3)$ & 255.6 & 245.0 & 260.9 & 246.0  \\ 
        $B_1(2)$ & 258.7 & 252.2 & 259.2 & 253.6  \\  \hline \hline
    \end{tabular}
    \label{tab:optical-modes}
\end{table*}

After diagonalizing the dynamical matrix, we extracted the sound velocities $v_{i,\alpha}$ from the long-wavelength dispersion relation $\omega_{\alpha}^{2} = \sum_{i} v_{i,\alpha}^{2} |\mathbf{q}_{i}|^{2}$, where $i$ labels the propagation direction and $\alpha$ denotes the polarization mode. Using Eqs.~\eqref{eq:sound-velocity-1st}--\eqref{eq:sound-velocity-end}, we derived expressions for the six independent elastic moduli in terms of the sound velocities along high-symmetry directions. The elastic coefficients for uniaxial compression along $x$ and $z$ axes are given by the standard relations
\begin{align}
C_{11} = \rho v^2_{x,l},
\qquad
C_{33} = \rho v^2_{z,l},
\end{align}
where the subindex $l$ denotes the longitudinal mode. The off-diagonal elastic constants associated with the Poisson-type couplings on the $x$–$y$ and $x$–$z$ planes, respectively, read
\begin{align}
&C_{12} = \rho \left( v^2_{x,l} - 2 v^2_{xy,t_1} \right), \\
&C_{13} = \rho \sqrt{\left(v^2_{xz,l} - v^2_{xz,t_1}\right)^{2} - \frac{1}{4}\left(v^2_{x,l} - v^2_{z,l}\right)^{2}} ,
\end{align}
where the subindex $t_1$ denotes the lowest transverse mode. Finally, the shear moduli associated with shear distortions on the $x$–$z$ and $x$–$y$ planes are given by
\begin{align}
&C_{44} = \rho \frac{v^2_{z,t_1} + v^2_{x,t_1} + v^2_{xy,t_2}}{3}, \\
&C_{66} = \rho v^2_{x,t_2},
\end{align}
respectively, where the subindex $t_2$ denotes the highest transverse mode. For $C_{44}$, we defined an average over three transverse sound velocities, which helps mitigate the dispersion observed in the numerical values extracted from the phonon spectrum.

As a minimal requirement, the elastic coefficients $C_{ij}$ must satisfy the following mechanical stability criteria~\cite{Wu_2007}:
\begin{align}
    &C_{11}, C_{33}, C_{44}, C_{66} >0  \label{eq:stability-1} \\
    &C_{11} > C_{12} , \; \quad C_{11} + C_{33} > 2 C_{13} \label{eq:stability-2} \\
    &2(C_{11} + C_{12}) + C_{33} + 4C_{13} > 0 \label{eq:stability-3} .
\end{align}

\section{Results} \label{sec:results}

\subsection{Electronic bands and Fermi surface}

The calculated lattice parameters for the tetragonal conventional cell and the mass density are listed in Table~\ref{tab:cell-parameters}  for LDA, PBE and PBEsol functionals, together with values extracted from selected experimental \cite{Besara_2016, Boller_1963, Furuseth_1965, Laliberte_2020, Saini_1964, Willerstrom1984} and numerical \cite{Buckeridge_2016, Chang_2016, Grassano_2018, Huang_2015, Liu_2017, Naher_2021, Ullah_2023} references, including their corresponding standard deviations. Our results with PBE and PBEsol functionals are in excellent agreement with the experimental values, with relative deviations on the order of $0.5\%$. In contrast, slightly larger errors are obtained with the LDA functional. This small discrepancy is consistent with the well-known tendency of GGA functionals to overestimate lattice parameters~\cite{Chang_2016}. Fig.~\ref{fig:unit-cells-TaAs} displays the primitive (left) and conventional (right) unit cells, visualized using the \textit{XCrysDen} program~\cite{Kokalj_1999, Kokalj_2003}.

\begin{figure}
    \centering
    \includegraphics[scale=0.22]{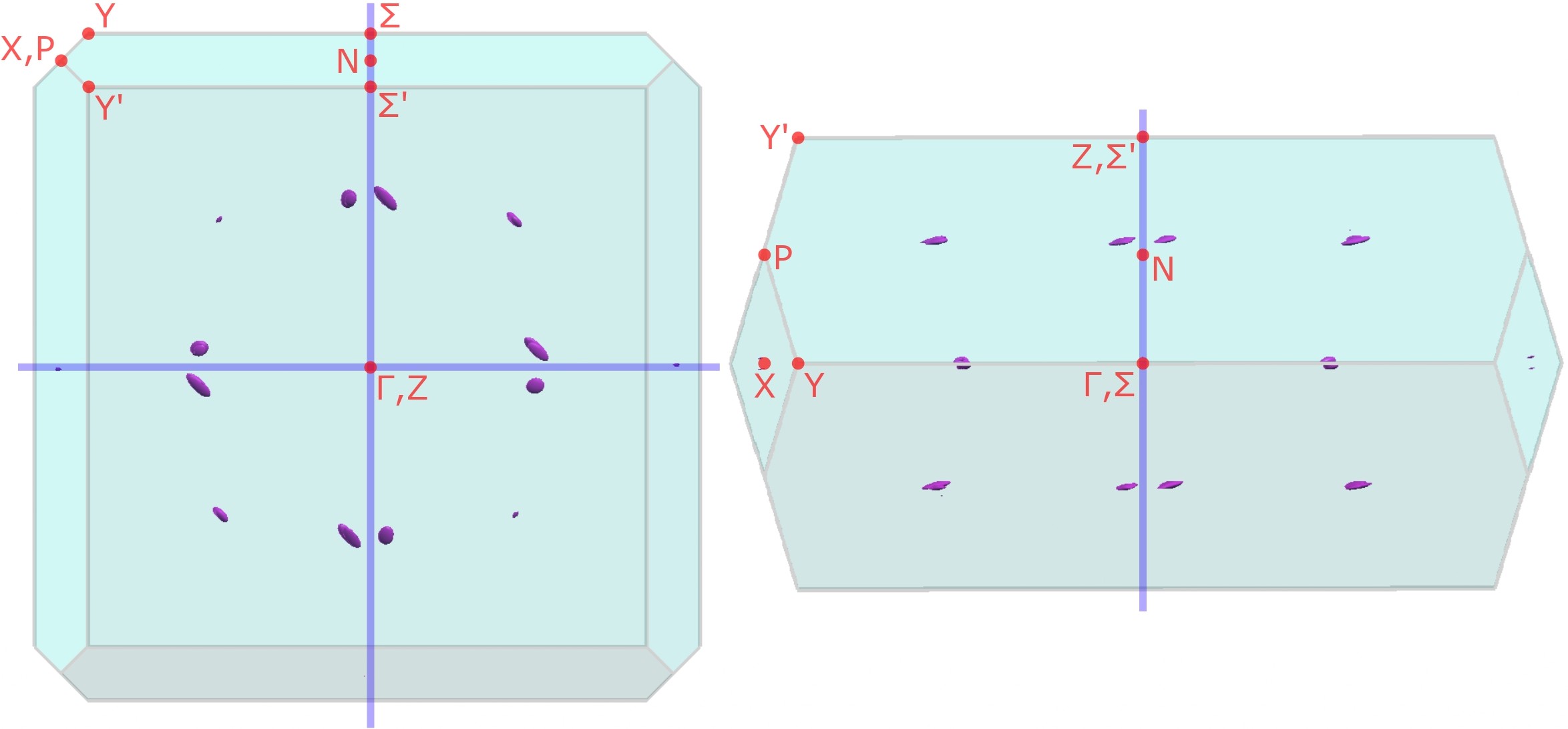}
    \caption{Fermi surface in the first Brillouin zone. Red dots denote the high-symmetry points, and blue lines indicate the mirror planes.}
    \label{fig:Fermi-surface}
\end{figure}

Using the lattice parameters shown in Table~\ref{tab:cell-parameters} and the atomic Wyckoff positions in fractional coordinates $(1/2, 0, 0.500)$ for Ta and $(1/2, 0, 0.082)$ for As, obtained from our fully relaxed cell geometry, we computed the electronic band structure shown in Fig.~\ref{fig:bands}, both with and without SOC. As expected, the band crossings near the $\Sigma$ and $\Sigma'$ points develop a gap once SOC is included~\cite{Grassano_2018, Huang_2015}. The resulting Fermi surface with SOC is presented in Fig.~\ref{fig:Fermi-surface}, where we observe that SOC gaps the band crossings located on the mirror planes, except for twelve pairs of Weyl nodes, each pair symmetrically separated by the corresponding mirror plane. There are only two independent pairs of Weyl nodes, since all remaining ones are generated from these two by applying the crystal symmetries. In addition, the Fermi surface features a trivial electron pocket along the $\Gamma$–$X$ segment, which appears four times due to the mirror symmetries of the Brillouin zone.

\begin{table*}
    \centering
    \caption{Sound velocities.}
    \begin{tabular}{c|c|c|c|c} \hline \hline
          & \multicolumn{2}{c|}{\textbf{This work}} & \textbf{Other DFT works} \cite{Buckeridge_2016, Chang_2016, Liu_2015, Liu_2017, Peng_2016} & \textbf{Expt. values} \cite{Laliberte_2020} \\ \hline
         \textbf{Direction} & \textbf{PBE} & \textbf{PBEsol} &  &  \\
         $v_{z,t_1}=v_{z,t_2}$ (km/s) & 2.57 & 2.68 & $2.85 \pm 0.33$ & - \\
         $v_{z,l}$ (km/s) & 4.54 & 4.64 & $4.30 \pm 0.26$ & - \\
         $v_{x,t_1}$ (km/s) & 2.95 & 2.99 & $3.11 \pm 0.16$ & - \\
         $v_{x,t_2}$ (km/s) & 3.90 & 3.93 & $3.80 \pm 0.19$ & - \\
         $v_{x,l}$ (km/s) & 4.73 & 4.92 & $4.83 \pm 0.12$ & - \\
         $v_{xy,t_1}$ (km/s) & 2.49 & 2.38 & $2.32 \pm 0.28$ & - \\
         $v_{xy,t_2}$ (km/s) & 2.86 & 2.91 & $2.75 \pm 0.29$ & 2.8 \\
         $v_{xy,l}$ (km/s) & 5.45 & 5.83 & $5.39 \pm 0.76$ & - \\
         $v_{xz,t_1}$ (km/s) & 2.55 & 2.71 & - & - \\
         $v_{xz,t_2}$ (km/s) & 3.63 & 3.61 & - & - \\
         $v_{xz,l}$ (km/s) & 4.78 & 4.94 & - & - \\ \hline \hline
    \end{tabular}
    \label{tab:sound-velocities}
\end{table*}

\subsection{Phonon dispersion and sound velocities}

\begin{figure}[b]
    \centering
    \includegraphics[scale=0.47]{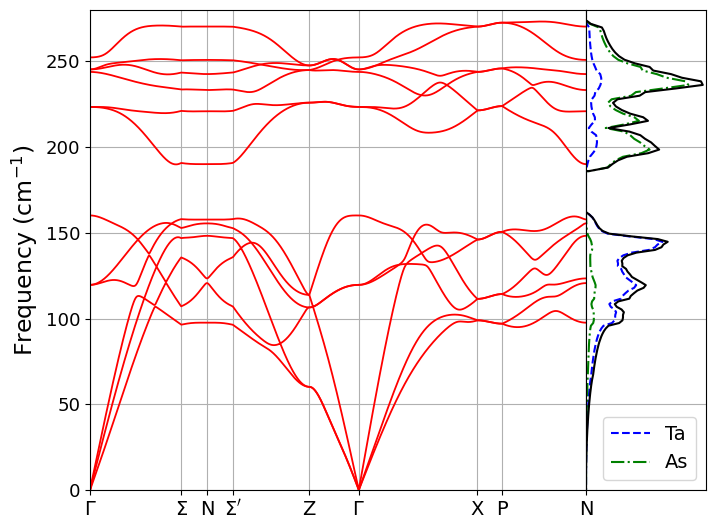}
    \caption{Phonon dispersion and DOS, based on the PBE.}
    \label{fig:phonons}
\end{figure}

The phonon spectrum and its corresponding dispersion play a central role in understanding the structural properties and transport phenomena in solids~\cite{Seol_2010, Munoz_2010, Munoz_2016}. The phonon spectrum and density of states (DOS) of TaAs computed in this work with the PBE are shown in Fig.~\ref{fig:phonons}, including SOC~\cite{Urru_PhysRevB.100.045115} which is physically relevant in Weyl semimetals, in contrast to most calculations reported in the literature~\cite{Buckeridge_2016, Chang_2016, Liu_2015, Liu_2017}. For methodological comparison, the phonon spectrum obtained with PBEsol is presented in Appendix~\ref{app:PBEsol}. Since the primitive unit cell contains four atoms, the system exhibits twelve phonon modes at each momentum, which split into three acoustic and nine optical branches near the $\Gamma$ point. The absence of imaginary frequencies in the calculated dispersion confirms the dynamical stability of the simulated structure.

The phonon dispersion exhibits an energy gap between the optical modes of $29.8\text{ cm}^{-1} hc \approx 3.69$ meV, where $h$ is Planck's constant and $c$ is the speed of light. As shown by the partial phonon DOS in Fig.~\ref{fig:phonons}, the vibrational degrees of freedom of Ta and As are largely decoupled, resulting in six low-energy modes dominated by Ta vibrations and six high-energy modes dominated by As vibrations. This energy gap can be attributed to the large mass difference between the two atomic species, together with the layered crystal structure of TaAs.

At the $\Gamma$ point, the optical frequencies listed in Table~\ref{tab:optical-modes}, together with the experimental values extracted from Ref.~\cite{Liu_2015}, and the averaged numerical data from Refs.~\cite{Liu_2015, Chang_2016}, are classified according to the irreducible representations of the $C_{4v}$ point group~\cite{Kroumova_2003}. The vibrations at the $\Gamma$ point decompose into the acoustic modes $A_1 + E$ and the optical branch $A_1 + 2B_1 + 3E$, where all modes are Raman active and only $A_1$ and $E$ are infrared active. The doubly degenerate $E$ modes correspond to in-plane vibrations within the $a$–$b$ plane, while the $A_1$ and $B_1$ modes describe vibrations along the $c$ axis~\cite{Liu_2015}. Our computed values are in very good agreement with previous numerical results reported in the literature~\cite{Chang_2016, Liu_2015}.

\begin{figure}[t]
    \centering
    \includegraphics[scale=0.57]{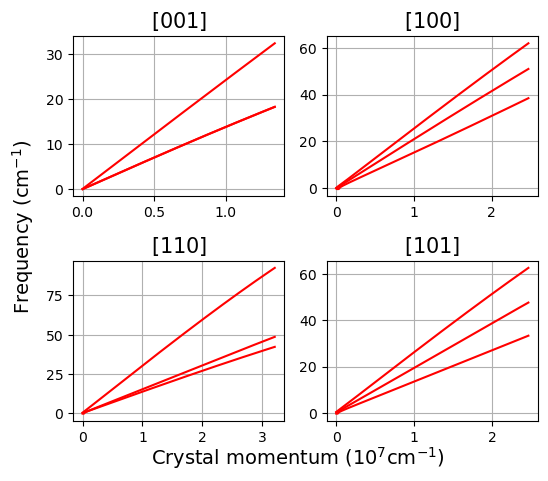}
    \caption{Acoustic modes extracted from the phonon spectrum computed with PBE.}
    \label{fig:sound-velocities}
\end{figure}

The acoustic modes along four different high-symmetry directions are displayed in Fig.~\ref{fig:sound-velocities}. From a linear fit of these low-energy modes, we extracted the corresponding sound velocities, listed in Table~\ref{tab:sound-velocities} for both PBE and PBEsol for methodological comparison, together with average values derived from Refs.~\cite{Buckeridge_2016, Chang_2016, Liu_2015, Liu_2017, Peng_2016} and their associated standard deviations. The polarization directions for each mode are specified in Appendix~\ref{appx:sound-velocities}. Our results show very good agreement with the literature, with all relative deviations below $10^{-1}$ with respect to the average value. Furthermore, our calculated value of $v_{xy,t_2} = 2.86$~km/s is in excellent agreement with Ref.~\cite{Laliberte_2020}, which reports an experimentally measured value of $v_{t_2}$ along the direction $[110]$ of $2.8$~km/s (see Table~\ref{tab:sound-velocities}).

\begin{table*}
    \centering
    \caption{Elastic moduli.}
    \begin{tabular}{c|c|cc|c} \hline \hline
         & \textbf{PBEsol} & \multicolumn{2}{c|}{\textbf{PBE}} & \quad \textbf{Other DFT works} \cite{Buckeridge_2016, Liu_2017, Naher_2021, Ullah_2023} \quad \\ \hline
         &  & \quad \textbf{Continuum model} \quad & \quad \textbf{DFT calculation} \quad &  \\ 
        $C_{11}$ (GPa) & 302.17 & 272.26 & 309.98 & $299.6 \pm 14.5$ \\
        $C_{33}$ (GPa) & 268.19 & 250.53 & 264.95 & $248.8 \pm 11.9$ \\
        $C_{12}$ (GPa) & 160.74 & 122.08 & 143.17 & $160.8 \pm 13.4$ \\
        $C_{13}$ (GPa) & 109.34 & 102.76 & 111.38 & $124.8 \pm 11.9$ \\
        $C_{44}$ (GPa) & 102.44 & 95.00  & -      & $89.6 \pm 4.8$ \\
        $C_{66}$ (GPa) & 193.88 & 184.60  & -      & $184.8 \pm 14.4$ \\ \hline \hline
    \end{tabular}
    \label{tab:elastic-moduli}
\end{table*}

\subsection{Elastic moduli}

\begin{figure}[b]
    \centering
    \includegraphics[scale=0.39]{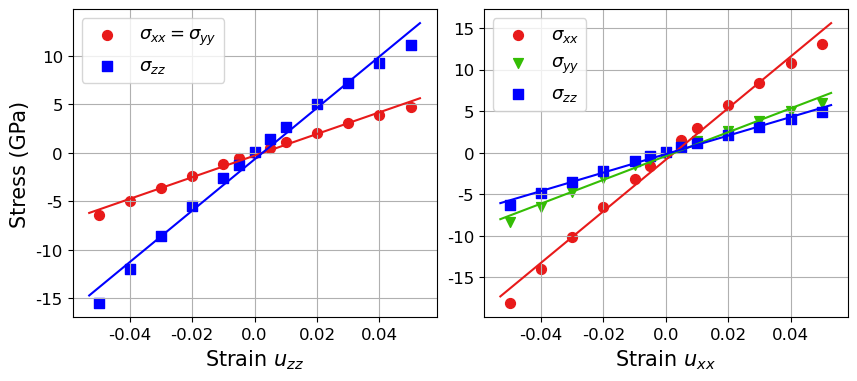}
    \caption{Stress as a function of the applied strain.}
    \label{fig:stress-strain}
\end{figure}

Our computed values for the elastic moduli, obtained both from the DFT-based strain–stress method and from combining the phonon spectrum with our continuum model, are summarized in Table~\ref{tab:elastic-moduli} including our results using PBE and PBEsol for methodological comparison, alongside averaged values extracted from Refs.~\cite{Buckeridge_2016, Liu_2017, Naher_2021, Ullah_2023} and their associated standard deviations. The strain–stress data used to compute the DFT-based elastic moduli are shown in Fig.~\ref{fig:stress-strain}. The three sets of values listed in Table \ref{tab:elastic-moduli} are fully consistent with the mechanical stability criteria described in Eqs.~\eqref{eq:stability-1}-\eqref{eq:stability-3}. The largest discrepancies in the continuum-model results occur for the coefficients $C_{12}$ and $C_{13}$, both with deviations of about $20\%$. In contrast, the relative deviations of all other elastic moduli remain below $10\%$, and in particular those of $C_{33}$ and $C_{66}$ are below $1\%$. To the best of our knowledge, there are no experimental measurements of these elastic coefficients for TaAs reported in the literature, so our comparison relays only on the DFT calculations previously reported by other groups~\cite{Buckeridge_2016, Liu_2017, Naher_2021, Ullah_2023}.

Using the elastic coefficients listed in Table \ref{tab:elastic-moduli}, we compute the Voigt and Reuss bounds for the bulk and shear moduli \cite{Hill_1952}. Introducing the auxiliary quantity $M = C_{11} + C_{12} + 2C_{33} - 4C_{13}$, the Voigt bounds for the bulk and shear moduli are given by \cite{Wu_2007}
\begin{align}
&B_V = \frac{1}{9}\left[2(C_{11} + C_{12}) + C_{33} + 4C_{13}\right] \\
&G_V = \frac{1}{30}(M + 3C_{11} - 3C_{12} + 12C_{44} + 6C_{66}) .
\end{align}
Defining $C^2 = (C_{11} + C_{12})C_{33} - 2C_{13}^2$, the corresponding Reuss bounds read
\begin{align}
&B_R = C^2/M \\
&G_R = 15/\left( \frac{18B_V}{C^2} + \frac{6}{C_{11} - C_{12}} + \frac{6}{C_{44}} + \frac{3}{C_{66}} \right) .
\end{align}
The Hill averages $B$ and $G$ are obtained as the arithmetic mean of the corresponding Voigt and Reuss bounds. Using these averages the Young's modulus $E$ and the Poisson's ratio $\nu$ can be evaluated according to \cite{Wu_2007}
\begin{align}
    E = 9BG/(3B + G) , \quad  \nu = (3B - E)/6B .
\end{align}
The resulting values are reported in Table~\ref{tab:derived-elastic-parameters}, along with the minimum, maximum, and average values derived from literature data for the elastic constants $C_{ij}$ extracted from Refs.~\cite{Buckeridge_2016, Liu_2017, Naher_2021, Ullah_2023}, which are compiled in Table~\ref{tab:elastic-moduli}.

Following Ref.~\cite{Pugh_1954}, the ratio $B/G$ quantifies the malleability of a material, with low values corresponding to brittle behavior and high values to malleable behavior, respectively. Using the commonly adopted threshold value of $1.75$ to distinguish between the two regimes~\cite{Liu_2017}, our calculated $B/G$, reported in Table~\ref{tab:derived-elastic-parameters} for both PBE and PBEsol, indicates a brittle behavior for TaAs, comparable to Zn ($B/G = 1.59$)~\cite{Pugh_1954}. In contrast, the averaged value reported in the literature lies in the malleable regime, with values comparable to those of magnesium ($B/G = 1.95$) and La ($B/G = 1.9$).

Poisson's ratio provides information about the characteristic of the bonding forces in the crystal~\cite{Chu_1995}. The lower and upper limits for central inter-atomic force solids are 0.25 and 0.5, and our calculated value is compared with other estimations reported in the literature in Table~\ref{tab:derived-elastic-parameters}, where the range and the average are displayed. According to our calculated $\nu = 0.497$, TaAs exhibits a Poisson's ratio even higher than Pt ($\nu = 0.44$), Pb ($\nu = 0.44$) and Ta ($\nu = 0.45$) \cite{Pugh_1954}.

Ref.~\cite{Naher_2021} applied a similar elastic model, but their procedure involves the direct computation of elastic constants $C_{ij}$ via DFT, with sound velocities calculated from these values. In contrast, our method does the converse, i.e. we compute the sound velocities from DFTP (including SOC), and then calculate the elastic parameters from the analytical equations obtained from the continuum elastic model described in Appendix~\ref{appx:sound-velocities}. Therefore, both methods will clearly differ in their predictions, and ultimately experimental measurements are required to assess the accuracy of each one.

\begin{table*}
    \centering
    \caption{Bulk, shear and Young's moduli and Poisson's ratio of TaAs. The column for other DFT works includes range (in parenthesis) and average values.}
    \begin{tabular}{c|c|c|c} \hline \hline
         &  \textbf{PBE}  &  \textbf{PBEsol}  &  \textbf{Other DFT works} \cite{Buckeridge_2016, Liu_2017, Naher_2021, Ullah_2023}  \\ \hline
        $B_V$ (GPa) & 161.14 & 181.26 & [169.63 - 198.81] 185.42 \\
        $B_R$ (GPa) & 160.37 & 178.381 & [166.29 - 194.91] 181.77 \\
        $B$ (GPa) & 160.75 & 179.82 & [167.96 - 196.86] 183.60 \\
        $G_V$ (GPa) & 106.08 & 112.63 & [96.70 - 107.04] 101.97 \\
        $G_R$ (GPa) & 95.72 & 100.66 & [87.06 - 94.40] 90.12 \\
        $G$ (GPa) & 100.90 & 106.64 & [91.88 - 100.72] 96.05 \\
        $B/G$ & 1.593 & 1.686 & [1.828 - 2.008]\,1.912 \\
        $E$ (GPa) & 250.33 & 267.12 & [233.13 - 256.65] 245.35 \\
        $\nu$ & 0.497 & 0.497 & [0.497 - 0.498] 0.4975 \\ \hline \hline 
    \end{tabular}
    \label{tab:derived-elastic-parameters}
\end{table*}

\section{Conclusion}  \label{sec:conclusion}

In this work we have studied the phonon and elastic properties of the Weyl semimetal TaAs by combining DFPT with an elastic continuum model. The phonon spectrum, computed including SOC, shows a clear separation between low-energy Ta modes and high-energy As modes, giving rise to a phonon band gap between optical branches. This behavior is naturally explained by the large mass difference between Ta and As and is further supported by the partial phonon DOS.

From the acoustic phonon branches near the $\Gamma$ point, we extracted sound velocities along several high-symmetry directions and used them to determine the independent elastic moduli within an elastic continuum framework. The resulting elastic coefficients are in good overall agreement with previous numerical studies, with the largest deviations appearing in $C_{12}$ and $C_{13}$. All obtained elastic moduli satisfy the mechanical stability criteria for tetragonal crystals, confirming the mechanical stability of TaAs within the linear elastic regime.

Using the calculated elastic constants, we further derived macroscopic mechanical parameters, including the Voigt–Reuss–Hill averages of the bulk and shear moduli, as well as the Young’s modulus and Poisson’s ratio. These derived quantities are consistent with values reported in the literature and provide a coherent description of the elastic response of TaAs. The resulting $B/G$ ratio, computed independently from PBE and PBEsol phonon spectra, places TaAs close to the brittle–malleable boundary, highlighting the sensitivity of such descriptors to the underlying elastic constants. 

An important advantage of the present approach is that the combination of phonon calculations with an elastic continuum model avoids the explicit application of shear strain and, by construction, circumvents Pulay-stress–related issues. This eliminates the need for very large supercells or repeated structural relaxations required in conventional stress–strain calculations. On the other hand, since elastic coefficients are strictly defined in the elastic limit of the material, calculations based on acoustic phonon spectra ensure that this condition is fulfilled by construction. In contrast, the direct application of strain-stress relays on static deformations of the unit cell where the elastic regime must be carefully monitored. At the same time, the accuracy of this methodology relies on well-resolved acoustic phonon branches and can be affected by numerical dispersion in the extracted sound velocities, making it complementary to, rather than a replacement for, direct DFT-based elastic calculations. Since our proposed method exhibits some discrepancies with the predictions arising from the more standard strain-stress calculations, a direct test of the elastic model applied to a broader pool of materials will provide a more general comparison between both methods.

Overall, our results provide a consistent characterization of the phonon spectrum and elastic properties of TaAs and show that elastic constants can be efficiently obtained from lattice-dynamical information. This combined approach is particularly suitable for topological materials, where lattice degrees of freedom play an important role and conventional elastic calculations can be computationally demanding. Moreover, while direct experimental characterization of the elastic properties of TaAs and other Weyl semimetals are not yet available in the literature, an accurate determination of the lattice dynamics would be possible by Raman scattering~\cite{Liu_2015,Gao_JAP_2007,Mishra_PRB_2001,Ramos_ACSNANO_2023,Weinstein_PhysRevB_1973}, particularly second-order Raman~\cite{Carles_PhysRevB_1980,Gao_JAP_2007,Carvalho_2017,Hildebrandt_2023,Iatsunskyi_Optik_2026,Mishra_PRB_2001,SOURISSEAU_1991,Weber_PRB_1993,Weinstein_PhysRevB_1973} for acoustic modes, field-angle dependence~\cite{Laliberte_2020}, Brillouin-Mandelstam scattering~\cite{Guzman_ACS_2022,Bottani01012018,Kargar_APL_2018,Kargar_NP_2021,Wright_APL_2024,Wright_APL_2025}, X-ray diffraction~\cite{Lee_PRL_2022} and other techniques~\cite{Bottani01012018,Rigg_2014}. Therefore, our proposed method based on the accurate determination of the speed of sound by DFPT with SOC~\cite{Baroni_2001,Urru_PhysRevB.100.045115} to infer the elastic moduli from an analytical elastic model, then offers an alternative and direct route to compare with experiments.

\begin{acknowledgments}

E.M. Acknowledges financial support from ANID Fondecyt grant No 1230440. F.J.P. acknowledges financial support from ANID Beca Doctorado Nacional 2022 Grant No. 21221919. D.G. was supported by the ANID Fondecyt Postdoctorado 2025 Grant No. 3250768. 

\end{acknowledgments}

\appendix

\section{Calculation with PBEsol functional}
\label{app:PBEsol}

\begin{figure}
    \centering
    \includegraphics[scale=0.525]{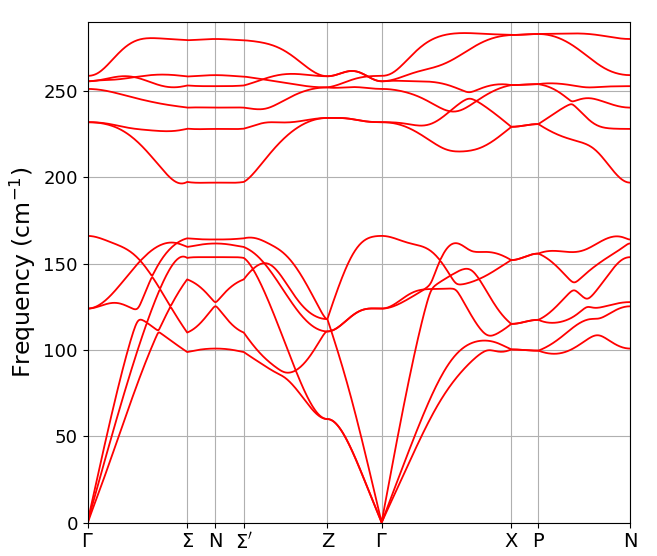}
    \caption{Phonon dispersion obtained with PBEsol functional.}
    \label{fig:phonons_PBEsol}
\end{figure}

In order to compare our results with a different functional, we repeated all phonon calculations and the corresponding derivation of the elastic coefficients using the PBEsol functional, that has been applied to compute structural and elastic propertied for various materials \cite{Haas_2009, Csonka_2009, Schmidt_2022}. For this calculation, we used the same parameters as with the PBE functional, \textit{i.e.} a 100 Ry plane-wave cutoff and a 12 $\times$ 12 $\times$ 12 Monkhorst--Pack k-point mesh for the unit cell relaxation, with stress and force component thresholds of $10^{-7}$ Ry/$a_B^3$ and $5 \times 10^{-7}$ Ry/$a_B$, respectively. In addition, phonon frequencies were obtained by performing a self-consistent electronic calculation on a $10 \times 10 \times 10$ $k$-point mesh with the same 100 Ry plane-wave cutoff, followed by DFPT including SOC on a $5 \times 5 \times 5$ $q$-point mesh. Sound velocities were extracted from the phonon spectrum shown in Fig.~\ref{fig:phonons_PBEsol} by performing a linear regression to the acoustic branches near the $\Gamma$ point. For the sake of comparison, the extracted values for the speed of sound obtained from this PBEsol generated phonon spectrum are reported in Table~\ref{tab:sound-velocities}, the computed elastic moduli in Table~\ref{tab:elastic-moduli}, and the elastic coefficients in Table~\ref{tab:derived-elastic-parameters}, respectively.

\section{Derivation of sound velocities in terms of elastic moduli} \label{appx:sound-velocities}

In the linear elastic regime, the general form for the free energy of a deformed crystal is~\cite{Landau_1986}
\begin{equation}
    F = \frac{1}{2} \sum_{iklm} \lambda_{iklm} u_{ik} u_{lm} \, , \label{eq:free-energy}
\end{equation}
where $u_{ij}$ and $\lambda_{ijkl}$ are the components of the strain and elasticity (or stiffness) tensors, respectively. Clearly from Eq.~\eqref{eq:free-energy}, the elasticity tensor must satisfy the symmetry properties $\lambda_{ijkl} = \lambda_{klij} = \lambda_{jikl} = \lambda_{ijlk}$. Deriving the stress tensor $\sigma_{ij} = \partial F/\partial u_{ij}$, the anisotropic Hooke's law is obtained:
\begin{equation}
    \sigma_{ij} = \sum_{k\ell} \lambda_{ijkl} u_{k\ell} .  \label{eq:stress-strain}
\end{equation}
Due to the mirror symmetries of the $C_{4v}$ point group, the elasticity tensor contains only six independent coefficients~\cite{Authier_2006}.

We rewrite the stress and strain tensors in terms of the vectors $\boldsymbol{\sigma} = (\sigma_{xx}, \sigma_{yy}, \sigma_{zz}, \sigma_{yz}, \sigma_{xz}, \sigma_{xy})$ and $\boldsymbol{u} = (u_{xx}, u_{yy}, u_{zz}, 2u_{yz}, 2u_{xz}, 2u_{xy})$. This Voigt mapping~\cite{Authier_2006} recasts the fourth-rank stiffness tensor into a $6 \times 6$ matrix with components $C_{ij} = C_{ji}$. For crystalline systems with $C_{4v}$ point group symmetry, this mapping reads
\begin{align}
    &C_{11} = C_{22} = \lambda_{1111} , \quad C_{33} = \lambda_{3333}, \\
    &C_{12} = C_{21} = \lambda_{1122} , \quad C_{13} = C_{31} = \lambda_{1133} , \\
    &C_{44} = \lambda_{1313} , \quad C_{66} = \lambda_{1212} ,
\end{align} 
and the other components vanish. In this notation, Eq.~\eqref{eq:stress-strain} becomes
\begin{equation}
\sigma_i = \sum_j C_{ij} u_j ,
\end{equation}
where $\sigma_i$ and $u_j$ denote the components of $\boldsymbol{\sigma}$ and $\boldsymbol{u}$, respectively.

\begin{widetext}

Starting from the stress–strain relation \eqref{eq:stress-strain}, the equations of motion for the displacement field $\Delta\boldsymbol{x}(\mathbf{x},t)$ follow as
\begin{equation}
    \rho \frac{d^2}{dt^2} \Delta{x}_i = \sum_j \frac{\partial \sigma_{ij}}{\partial x_j} = \frac{1}{2} \sum_{jkl} \lambda_{ijkl} \partial_j \partial_k \Delta{x}_l ,
\end{equation}
where $t$ is time and $x_i$ is the $i$-th component of an arbitrary point in the elastic medium.

The resulting equations of motion for each component of the displacement field $\Delta \boldsymbol{x} = (\Delta x, \Delta y, \Delta z)$ are
\begin{align}
    &\rho \frac{d^2}{dt^2} \Delta x = {\left(C_{11} \partial_{x}^{2} + C_{66} \partial_{y}^{2} + C_{44} \partial_{z}^{2}\right) \Delta x } + \left( C_{12} + C_{66} \right) \partial_{x} \partial_{y} \Delta y + \left( C_{13} + C_{44} \right) \partial_{x} \partial_{z} \Delta z , \label{eq:wave-equation-x} \\
    &\rho \frac{d^2}{dt^2} \Delta y = (C_{12} + C_{66}) \partial_{x} \partial_{y} \Delta x  + \left(C_{66} \partial_{x}^{2} + C_{11} \partial_{y}^{2} + C_{44} \partial_{z}^{2}\right) \Delta y + (C_{13} + C_{44}) \partial_{y} \partial_{z} \Delta z  , \label{eq:wave-equation-y} \\
    &\rho \frac{d^2}{dt^2} \Delta z = (C_{13} + C_{44}) \partial_{x} \partial_{z} \Delta x + (C_{13} + C_{44}) \partial_{y} \partial_{z} \Delta y + \left(C_{44}\left(\partial_{x}^{2}+\partial_{y}^{2}\right) + C_{33}^{2} \partial_{z}^{2}\right) \Delta z  , \label{eq:wave-equation-z}
\end{align}
where we used $\partial_i \equiv \partial/\partial x_i$ to denote the spatial derivatives along the direction $x_i$. Upon Fourier transforming the displacement field according to $\Delta\boldsymbol{x}(\mathbf{x},t) = \sum_{\mathbf{q},\omega} \Delta\boldsymbol{x}(\mathbf{q},\omega) e^{i(\mathbf{q} \cdot \mathbf{x}-\omega t)}$, with $\mathbf{q}$ the wave vector and $\omega$ the angular frequency, the wave equations reduce to the eigenvalue problem
\begin{equation}
    \left( \mathbf{M}(\mbf{q}) - \rho \omega \right) \Delta\boldsymbol{x}(\mathbf{q}, \omega) = 0 ,  \label{eq:dispersion-eigenvalues}
\end{equation}
where $\mathbf{M}(\mathbf{q})$ is the dynamical matrix, whose components are given by $M_{ij}(\mathbf{q}) = \sum_{kl} \lambda_{iklj} q_k q_l$. From Eqs.~\eqref{eq:wave-equation-x}-\eqref{eq:wave-equation-z}, the explicit form of $\mathbf{M}(\mathbf{q})$ for a body-centered tetragonal crystal, expressed in terms of the elastic coefficients $C_{ij}$, is 
\begin{equation} \label{eq:dynamical-matrix}
    \mathbf{M}(\mbf{q}) =
    \begin{pmatrix}
      C_{11} q_x^2 + C_{66}q_y^2 + C_{44}q_z^2 & (C_{12} + C_{66}) q_x q_y & (C_{13} + C_{44}) q_x q_z  \\
      (C_{12} + C_{66}) q_x q_y & C_{66}q_x^2 + C_{11} q_y^2 + C_{44} q_z^2 & (C_{13} + C_{44}) q_y q_z  \\
      (C_{13} + C_{44}) q_x q_z & (C_{13} + C_{44}) q_y q_z & C_{44}\left( q_x^2 + q_y^2 \right) + C_{33} q_z^2   
    \end{pmatrix} .
\end{equation}

Eq.~\eqref{eq:dispersion-eigenvalues} defines the dispersion relations $\rho \omega_{\alpha}(\mathbf{q})$ of the sound waves as eigenvalues of the dynamical matrix $\mathbf{M}(\mbf{q})$, where $\alpha \in \{l,t_1,t_2\}$ represents each of the three acoustic polarization modes. In terms of the sound velocities $v_{i,\alpha}^{2}$ along the direction $i$ and with polarization mode $\alpha$, the dispersion relations are expressed as $\omega_{\alpha}^{2} = \sum_{i} v_{i,\alpha}^{2} |\mathbf{q}_{i}|^{2}$. Thus, the sound velocities along high-symmetry directions are obtained by evaluating the corresponding unit vectors in Eq.~\eqref{eq:dynamical-matrix} and diagonalizing the resulting matrix. For propagation along $[001]$, the longitudinal and transverse modes satisfy
\begin{align}
&\rho v_{z,l}^2 = C_{33}, \quad 
\rho v_{z,t_1}^2 = \rho v_{z,t_2}^2 = C_{44},  \label{eq:sound-velocity-1st}
\end{align}
with polarizations along $[001]$, $[100]$, and $[010]$, respectively. For propagation along $[100]$ (or equivalently $[010]$), the velocity relations become
\begin{align}
&\rho v_{x,l}^2 = C_{11}, \quad 
\rho v_{x,t_1}^2 = C_{44}, \quad 
\rho v_{x,t_2}^2 = C_{66},
\end{align}
and the corresponding polarizations lie along $[100]$, $[001]$, and $[010]$. Along the diagonal direction $[110]$, we obtain
\begin{align}
\rho v_{xy,l}^2 = \frac{1}{2}\left( C_{11} + C_{12} \right) + C_{66}, \quad
\rho v_{xy,t_1}^2 = \frac{1}{2}\left( C_{11} - C_{12} \right), \quad
\rho v_{xy,t_2}^2 = C_{44},
\end{align}
with polarizations oriented along $[110]$, $[1\bar{1}0]$, and $[001]$. Finally, for propagation along $[101]$ (or equivalently $[011]$) the velocities satisfy
\begin{align}
\rho v_{xz,l}^2 = \frac{1}{4} \left( \Delta_1 + \Delta_2 \right), \quad
\rho v_{xz,t_1}^2 = \frac{1}{4} \left( \Delta_1 - \Delta_2 \right), \quad
\rho v_{xz,t_2}^2 = \frac{1}{2}\left( C_{44} + C_{66} \right),   \label{eq:sound-velocity-end}
\end{align}
where $\Delta_1 = C_{11} + C_{33} + 2C_{44}$ and $\Delta_2 = \sqrt{( C_{11} - C_{33} )^2 + 4( C_{13} + C_{44} )^2}$, and the polarizations lie along $[101]$, $[10\bar{1}]$, and $[010]$. Collectively, these relations form an overdetermined system for the elastic moduli $C_{11}$, $C_{33}$, $C_{12}$, $C_{13}$, $C_{44}$, and $C_{66}$ in terms of the sound velocities, so their resultant expressions are not unique.

\end{widetext}


\bibliography{ref.bib}

\end{document}